\documentclass[5p]{elsarticle}

\usepackage{bm}
\usepackage{graphicx}
\usepackage{color}
\usepackage{amsmath}
\usepackage{hyperref}
\usepackage{tikz}

\usetikzlibrary{decorations.pathreplacing}
\usepackage[absolute,overlay]{textpos}

\DeclareMathAlphabet\mathbfcal{OMS}{cmsy}{b}{n}

\journal{Physics Letters B}

\newcommand{\nhid}{N_\textrm{hid} }

\begin{document}

\begin{frontmatter}

\title{Machine learning the deuteron}

\author[surrey]{JWT Keeble}
\author[surrey]{A Rios\corref{mycorrespondingauthor}}
\cortext[mycorrespondingauthor]{Corresponding author}
\ead{a.rios@surrey.ac.uk}

\address[surrey]{Department of Physics, Faculty of Engineering and Physical Sciences, University of Surrey, Guildford, Surrey GU2 7XH, United Kingdom}

\begin{abstract}
We use machine learning techniques to solve the nuclear two-body bound state problem, the deuteron. We use a minimal one-layer, feed-forward neural network to represent the deuteron $S-$ and $D-$state wavefunction in momentum space, and solve the problem variationally using ready-made machine learning tools. We benchmark our results with exact diagonalisation solutions. We find that a network with $6$ hidden nodes (or $24$ parameters) can provide a faithful representation of the ground state wavefunction, with a binding energy that is within $0.1 \, \%$ of exact results. This exploratory proof-of-principle simulation 
may provide insight for future potential solutions of the nuclear many-body problem using variational artificial neural network techniques.
\end{abstract}


\end{frontmatter}

\def\layersep{2.5cm}

\section{Introduction}

Machine learning (ML) techniques are ubiquitous within and outside the scientific domain. They are used in a variety of contexts and can be exploited to classify information; to compress it; to interpolate or extrapolate data, and to solve a variety of optimisation problems  \cite{MacKay}.  
In physics, artificial neural networks (ANNs) have been extensively used in the past to analyse data, particularly in particle physics experiments and theory \cite{Feindt2006,Ball2010}. In nuclear physics, early applications of ANNs to nuclear systematics \cite{Gernoth1993,Gazula1992} have been recently extended to exotic mass domains \cite{Utama2016}, fission yields \cite{Wang2019}, $\beta-$ and $\alpha-$decay half-lives \cite{Niu2019,Freitas2019} and nuclear deformation and spectroscopic properties \cite{Regnier2019}. In \emph{ab initio} nuclear structure theory, ANNs can be used to extrapolate results of otherwise costly first-principles calculations from restricted model spaces \cite{Negoita2018,Negoita2019,Jiang2019}. 

A more recent development of ML techniques is their application to solve specific physics problems in the quantum domain \cite{Dunjko2018,Mehta2019,Carleo2019}. In particular, a series of recent ML applications have shown promising results in the solution of quantum many-body problems from first principles. The pioneering application of Ref.~\cite{Carleo2017} in spin systems used a restricted Boltzmann machine as a wavefunction ansatz. These simulations give access to both the ground state and the dynamics of systems with different dimensions, and extensions to excited states have also been formulated \cite{Choo2018}. The solution of discrete \cite{Saito2017} and real space \cite{Saito2018} many-body bosonic systems followed shortly after. More sophisticated techniques based on deep neural networks have been recently developed to tackle realistic quantum chemistry problems \cite{Pfau2019,Hermann2019,Choo2020}. In all these cases, the problem is set up as a variational one, and the solution is fully \emph{ab initio}. While we were preparing this manuscript, the preprint in Ref.~\cite{Adams2020} reported results for few-body nuclei similar in spirit to what we report here.

There are two key reasons that make ANNs particularly attractive in the quantum many-body domain. First, ANNs can encapsulate and compress information. If this compression is efficient enough, the complex content of many-body wavefunctions may be codified into manageable, specifically tailored and, possibly, deep ANNs \cite{Gao2017}. Second, ML techniques are particularly suited to solve optimisation problems. In a physics setting, with the energy as a cost function, these can be easily mapped into variational problems. The expectation is that these variational artificial neural network (VANNs) are superior to traditional trial wavefunctions, due to their ability to express features flexibly and efficiently. 

By providing direct access to the many-body wavefunction, ML techniques open a series of interesting possibilities to find nuclear ground states, operator expectation values and dynamics. 
Whether or not one can actually implement VANN algorithms efficiently in nuclear many-body systems is at present an open question. 
Here, we present a proof-of-principle calculation of a nuclear system, the deuteron, using ready-made, available ML resources. The deuteron is a natural starting point to explore the feasibility of \emph{ab initio} methods \cite{Dumitrescu2018}.
%, which already includes some of the difficulties encountered in heavier systems at a much lower computational cost .
While this is far from being a relevant many-body application, it allows for an exploratory analysis of the quality of ANN ans\"atze to the deuteron wavefunction. 
%It also demonstrates that a variational setting is useful in this context. 
%In solving the deuteron using VANNs, we identify some issues that may be representative for future developments of fully-fledged nuclear many-body ML solvers. 

\section{Methods}
\label{sec:methods}

\begin{figure}
\centering
\begin{tikzpicture}[shorten >=1pt,->,draw=black!50, node distance=\layersep,scale=0.75]
    \tikzstyle{every pin edge}=[<-,shorten <=1pt]
    \tikzstyle{neuron}=[circle,line width=5mm,fill=black!25,minimum size=17pt,inner sep=0pt]
    \tikzstyle{input neuron}=[neuron, fill=green!50];
    \tikzstyle{bias neuron}=[neuron, fill=green!50];
    \tikzstyle{output neuron}=[neuron, fill=red!50];
    \tikzstyle{hidden neuron}=[neuron, fill=blue!50];
    \tikzstyle{annot} = [text width=4em, text centered]

    % Draw the input layer nodes
    \foreach \name / \y in {1,...,1}
    % This is the same as writing \foreach \name / \y in {1/1,2/2,3/3,4/4}
        \node[input neuron, pin=left:$q$] (I-\name) at (0,-2) {};
     
    % Draw the input bias node 
	\foreach \name / \y in {1,...,1}
    % This is the same as writing \foreach \name / \y in {1/1,2/2,3/3,4/4}
        \node[bias neuron, pin=left: $\mathbf{b}$] (B-\name) at (0.5,-4) {};
        
    % Draw the hidden layer nodes
    \foreach \name / \y in {1,...,4}
        \path[yshift=0.5cm]
            node[hidden neuron] (H-\name) at (\layersep,-\y cm) {};
	
	% Draw the output layer nodes
	\foreach \name / \y in {2,4}
    	\path[yshift=0.5cm]
            node[output neuron] (O-\name) at (5,-\y cm) {};

	% Label the output nodes
	\node[output neuron, pin={[pin edge={->}]right:$\vert \psi_{ANN}^{S} \rangle$}] (O-2) at (5,-1.5) {};
	\node[output neuron, pin={[pin edge={->}]right:$\vert \psi_{ANN}^{D} \rangle$}] (0-4) at (5,-3.5) {};

    % Annotate weights
    \node (w1) at (1,0) {$\mathbfcal{W}^{(1)}$};
    \node (w2) at (4,0) {$\mathbfcal{W}^{(2)}$};

    % Connect every node in the input layer with every node in the
    % hidden layer.
    \foreach \source in {1,...,1}
        \foreach \dest in {1,...,4}
            \path (I-\source) edge (H-\dest);

    \foreach \source in {1,...,1}
        \foreach \dest in {1,...,4}
            \path (B-\source) edge (H-\dest);

    % Connect every node in the hidden layer with the output layer
    \foreach \source in {1,...,4}
    	\foreach \dest in {2,4}
        	\path (H-\source) edge (O-\dest);

    % Annotate the layers
    \node[annot,node distance=1cm] (hl) at (\layersep, 1cm) {\color{black} Hidden layer};
    \node[annot,left of=hl] {\color{black}Input layer};
    \node[annot,right of=hl] {\color{black}Output layer};
\end{tikzpicture}
\caption{\label{fig:NN}  ANN architecture used in this work. The input is a single value of momentum, $q$, and the wavefunctions are modelled in terms of a minimal single-layer network.  In the example above, the number of hidden nodes is $\nhid=4$. The ANN has two outputs, one for the $S$ and one for the $D$ state.
}  
\end{figure}
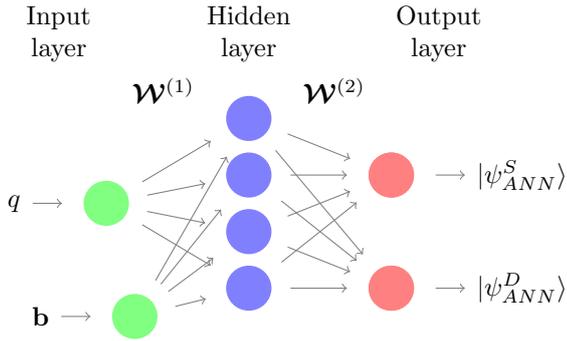

Our solution for the deuteron is variational. We set up a \emph{minimal} trial wavefunction. Our ANN has a single input node: a value of relative momentum, $q$, between the neutron and the proton in the deuteron. The ANN has two output nodes, one for the $L=0$ ($S$) and one for the $L=2$ ($D$) state. In between, we set up a single layer with  $\nhid$ hidden nodes. The architecture of the network is shown in Fig.~\ref{fig:NN}, which translates mathematically into a wavefunction ansatz
\begin{align}
    \psi_{\textrm{ANN}}^L (q) =  \sum_{i=1}^{N_{hid}} \mathcal{W}^{(2)}_{i,L} \, \sigma \left( \mathcal{W}^{(1)}_{i} q + b_{i} \right) \, ,
    \label{eq:wfANN}
\end{align}
where $\sigma(x)$ represents a non-linear activation function. 
The weights $\mathbfcal{W}^{(1)}$ connect the input relative momentum, $q$, to a hidden layer, whereas $\mathbfcal{W}^{(2)}$ connects the hidden layer to the two outputs. We also use a bias between the input and the hidden layer, $\mathbf b$.   We use bold notation $\mathbfcal{W}^{(1)}$ to denote the full weight (or bias) vectors, as opposed to the vector components $\mathcal{W}^{(1)}_i$. The concatenation of all weights and biases is denoted by $\mathcal{W}=\left\{ {\bf b}, \mathbfcal{W}^{(1)}, \mathbfcal{W}^{(2)} \right\}$. For a given number of hidden layer nodes $\nhid$, there are a total of $4 \nhid$ parameters  in the trial ANN wavefunction.

We use both a sigmoid and a softplus activation function $\sigma(x)$ in our ansatz. The two functions are  continuous and differentiable, and softplus is less prone to be affected by the vanishing gradient problem \cite{Hochreiter2001}. The output layer is a weighted linear sum of the values of the hidden nodes, and provides arbitrary admixtures of  the $S-$ and $D-$ states,
$\psi^{L=0,2}_\textrm{ANN}$. Dedicating a single layer to each of the two states would result in an increase of the number of parameters, departing from the minimal spirit of our approach. 
%In this sense, our ansatz ANN is a minimal representation of the deuteron wavefunction. 

The parameters $\mathcal{W}$ are used as variational parameters in a minimisation problem for the energy,
\begin{align}
E^{\mathcal{W}} = \frac{ \left \langle \Psi_\textrm{ANN}^{\mathcal{W}} \right | \hat H  \left | \Psi_\textrm{ANN}^{\mathcal{W}} \right \rangle}{ \left \langle \Psi_\textrm{ANN}^{\mathcal{W}} \right | \left. \Psi_\textrm{ANN}^{\mathcal{W}} \right \rangle} \, .
\label{eq:enerANN}
\end{align}
We solve the problem explicitly in momentum space \cite{Bogner2006,Bogner2006b,Anderson2010}. This is unlike previous VANN applications \cite{Saito2018,Pfau2019,Hermann2019,Choo2020},
but helpful for three practical reasons. First, in momentum space the kinetic term in the Hamiltonian of Eq.~(\ref{eq:enerANN}) is a continuous function. In contrast, in real space, the kinetic term would involve numerically costly derivatives on the ANN wavefunctions. Second, for the deuteron, the separation between centre-of-mass and relative motion can be implemented straightforwardly. The centre-of-mass coordinate can be ignored and the problem is solved as an effective one-body  Schr\"odinger equation in relative momentum, $q$.
 %With this, we avoid any potential sign problem. 
Third, a momentum space approach allows us to employ directly the numerical routines associated to the N3LO Entem-Machleidt nucleon-nucleon force, our interaction of choice \cite{Entem2003}. We have tested the method with other momentum-space potentials, and have found similar levels of agreement with the corresponding benchmarks. 

We use the same momentum quadrature in all our integrals. In the many-body case, these integrals may be more efficiently performed using Monte Carlo techniques \cite{Saito2018}. For the one-dimensional integrals associated to the deuteron, we estimate that a large number of Monte Carlo samples of order $>10^5$ is needed to get an accurate prediction for the binding energy. 
We instead use $N_k=64$ points in a Gauss-Legendre quadrature, and use a tangential change of variables to extend the integration range from $0$ to $k_\textrm{max}=500$ fm$^{-1}$.
%A change of variables involving a tangent is performed on a Gauss-Legendre quadrature of $N_k=64$ points to extend the integration range from $0$ to $k_\textrm{max}=500$ fm$^{1}$. 
This approach provides a dense mesh at low momenta, while sparsely covering the high-momentum region (only $7$ mesh points lie beyond $k=5$ fm$^{-1}$). We use the same quadrature to solve the exact ground state eigenvalue problem, to set a benchmark for the VANN solution and find an ``exact" ground state energy,  $E_\textrm{GS}=-2.2267$ MeV. 

The choice of a continuous momentum basis, as opposed to a discrete basis, is further motivated by an important result on ANNs. 
The Universal Approximation Theorem guarantees that a network with a single layer  provides a faithful representation of any continuous function within a given domain, provided $\nhid$ is large enough \cite{Cybenko1989,Hornik1991}. In this sense, working in continuous momentum space, rather than in a discrete basis, may be advantageous. One naively expects that ANNs should mimic the shape of any wavefunction, if given enough hidden nodes to do so. We note that perfect agreement between input and output is likely to require a local cost function, to penalise differences throughout momentum space. This is not necessarily the case here, where we use a global (integrated) energy cost function. 

%We stress here that the ANN ansatz can explore arbitrary shapes in momentum space in an unbiased way. The ansatz is thus not constrained by any prescriptive analytical shape, like traditional trial wavefunctions would be. This explains why the variational problem works here with a single wavefunction. In contrast, variational calculations with trial wavefunctions with $2-6$ parameters may not able to bind the deuteron \cite{Anderson2010}. 

\begin{figure}
\begin{center}
\includegraphics[width=\linewidth]{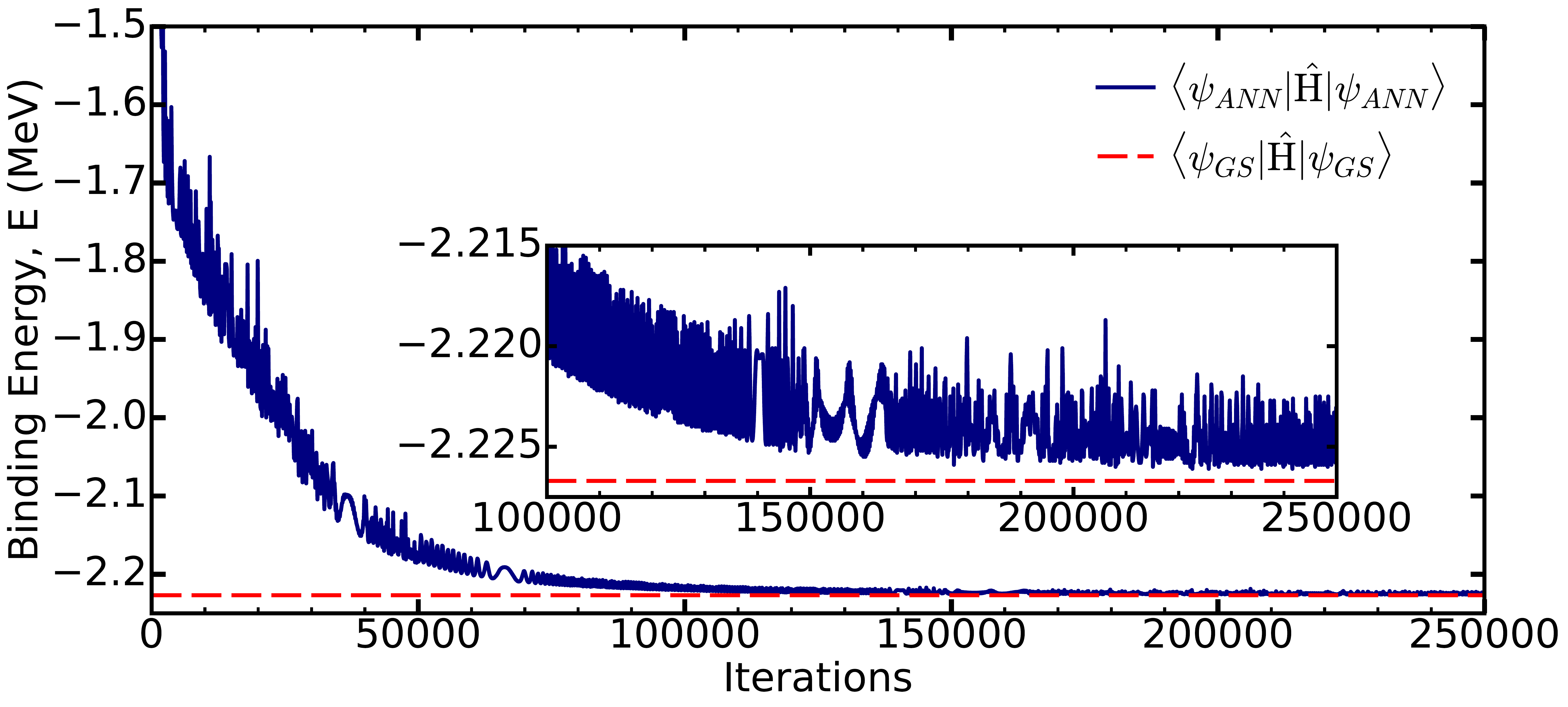}
\caption{\label{fig:enemin} Deuteron binding energy as a function of iteration number for a network with $\nhid=10$ and a softplus activation function. The energy cost function is minimised using RMSprop (see \ref{app:RMSprop} for details). } 
\end{center}
\end{figure}

We solve the variational problem in three different steps, implemented using the ready-made PyTorch framework \cite{PyTorch}. First, we initialise the network using random weight values. We sample from uniform distributions with $\mathbfcal{W}^{(1)} \in [-1,0)$, ${\bf b} \in [-1,1)$ and $\mathbfcal{W}^{(2)} \in [0,1)$. This differs from the traditional Xavier initialisation scheme, which has a poor performance in this problem \cite{Glorot2010}. After this random initialisation, the  wavefunctions are featureless and have no bearing to physical ones. In a second step, we therefore follow Ref.~\cite{Saito2018} and train the ANN to reproduce physically inspired, but arbitrary, target wavefunctions for each of the two states. 
%These target wavefunctions should be a closer starting point to the variational problem. 
We use a functional form $\psi_\textrm{targ}^{L}(q) \propto q^L e^{-\frac{\xi^2 q^2}{2} }$ with $\xi = 1.5$ fm, which provides target wavefunctions with  momentum space widths which are similar to the exact solutions.

We train the ANN wavefunction  to match the target wavefunction in a supervised manner. The cost function, $\mathcal{C}=\mathcal{C}^S+\mathcal{C}^D$, is the sum of the individual contributions for each state, $\mathcal{C}^L=(\mathcal{K}^L-1)^2$, where we introduce the overlap 
\begin{align}
\mathcal{K}^L &=\frac{ \left \langle \psi_\textrm{targ}^L \right | \left. \psi_\textrm{ANN}^L \right \rangle^2 }{
\left \langle \psi_\textrm{targ}^L \right | \left. \psi_\textrm{targ}^L \right \rangle
\left \langle \psi_\textrm{ANN}^L \right | \left. \psi_\textrm{ANN}^L \right \rangle}  
\label{eq:overlap}
\\ &=
\frac{[ \int_0^\infty d q \, q^2 \,  \psi_\textrm{targ}^L(q) \psi_\textrm{ANN}^L (q) ]^2}{
 \int_0^\infty d q \, q^2 \,  \psi_\textrm{targ}^L(q) \psi_\textrm{targ}^L (q)  \int_0^\infty d q \, q^2 \,  \psi_\textrm{ANN}^L(q) \psi_\textrm{ANN}^L (q) }  \, . \nonumber
\end{align}
The RMSprop scheme is used to minimise $\mathcal{C}$ for $10^5$ iterations  \cite{Mehta2019,Hinton2012}. 
We provide more details about this scheme in~\ref{app:RMSprop}, and list here only the relevant hyperparameters: $\alpha = 10^{-2}$, $\beta = 0.9$ and $\epsilon = 10^{-8}$. 
The network calculates an unnormalised wavefunction for each partial wave. In the minimisation algorithm, the wavefunction normalization constants divide the learning rates. Because these normalization constants are typically larger than one, 
%the normalisation constant effectively modulates $\mathcal{W}^{(2)}$ by a factor of $\sqrt{\sum_{L=0,2} \langle \psi_\textrm{ANN}^{L} \vert \psi_\textrm{ANN}^{L} \rangle}$. 
unnormalized wavefunctions effectively reduce the learning rate during the minimisation process, allowing for a relatively large value of $\alpha$. 
After this initial training step, the resulting overlap is within $1-5 \%$ of the desired value of $\mathcal{K}^L=1$. 
The admixture of the $S-$ and the $D-$states is deliberately chosen to have an unphysically large value of $50 \, \%$. 

The third and final step is the actual variational energy minimisation. We let the network evolve to readjust the wavefunctions while minimising the energy. The initial large admixture between the two states does not hinder the convergence of the VANN.  
We use RMSprop again to minimise the energy cost function in Eq.~(\ref{eq:enerANN}), with the same hyperparameter set discussed above. A typical energy minimisation curve for the case with $\nhid=10$ and a softplus activation function is shown in Fig.~\ref{fig:enemin}.
Within the first few thousands of iterations (not shown in the Figure for clarity), the descent is fast and smooth and the network is able to bind the deuteron. After about $10,000$ iterations, fluctuations appear. This allows for the energy to be overshot at times, but the minimisation algorithm eventually corrects for that.  
At $50,000$ iterations, the binding energy is already within $10 \%$ of the benchmark value (dashed line). We stop our runs at $250,000$ iterations, where the binding energy is converged within fluctuations of the order of $2-3$ keV. 

\section{Results}
\label{sec:results}

\begin{figure}
\begin{center}
\includegraphics[width=\linewidth]{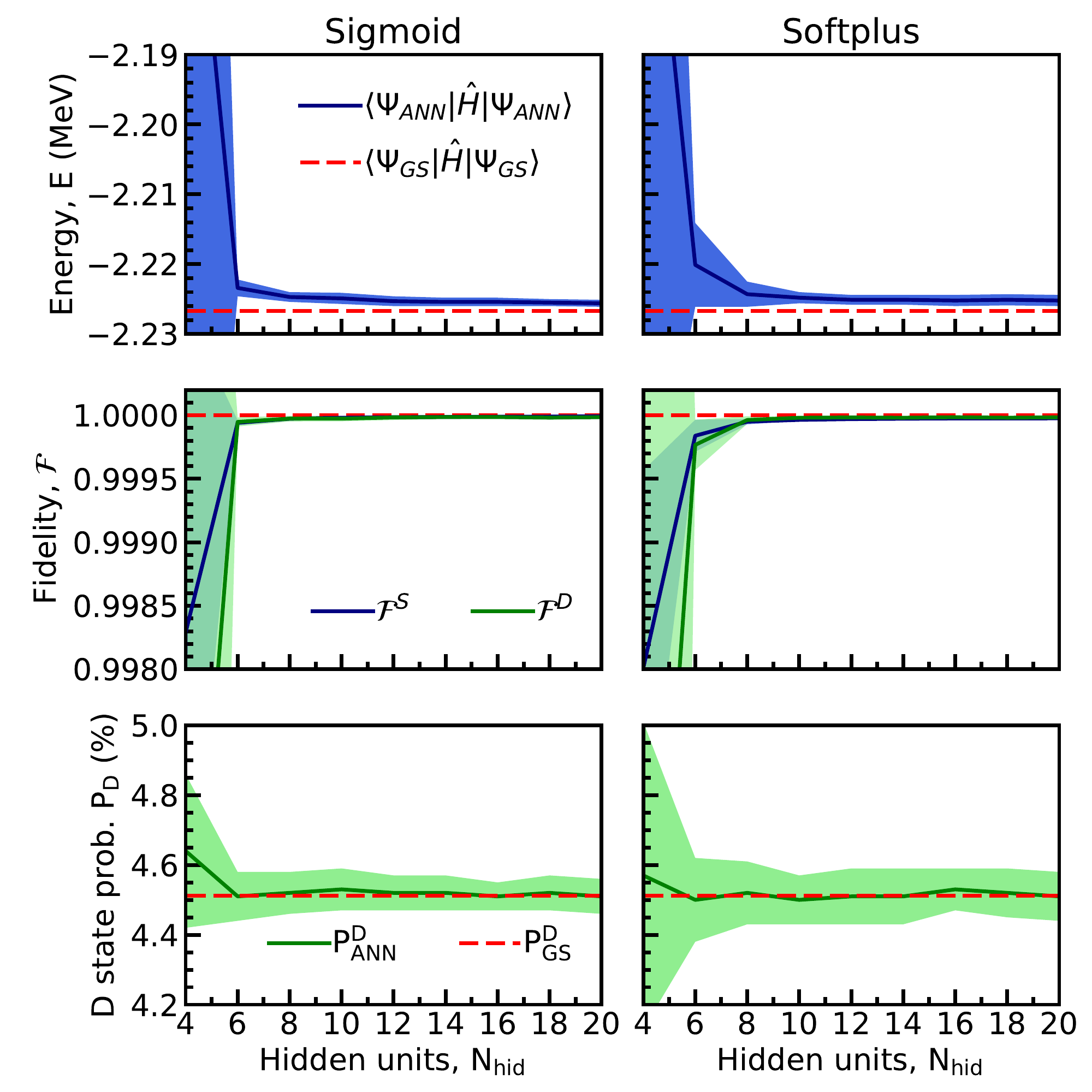}
\caption{\label{fig:energyNhid}  Binding energy of the deuteron (top panel), fidelities $\mathcal{F}^L$ (central) and $D-$state probability (bottom) as a function of the number of hidden layer nodes, $\nhid$. Lines (bands) are obtained from the average (standard deviation) of $50$ independent VANN runs. Horizontal (dashed) lines show the benchmark result. } 
\end{center}
\end{figure}

We explore the bias and variance of our minimal VANN model, particularly the out-of-sample error, in two different ways. First, we change the number of hidden layer nodes from $\nhid=2$ to $20$, in steps of $2$. An extended discussion up to $\nhid=100$ is presented in~\ref{app:variance}. This provides an idea of how model predictions change with an increase in the number of variational parameters. Second, we initialise the model, train it to target wavefunctions and minimise the energy with $50$ different random seed configurations. The results shown here are obtained as the means and standard deviations of these $50$ individual runs.  
This helps identify weight initialisation effects. 

\begin{table*}
 \caption{\label{table}
 VANN results for the fidelities $\mathcal{F}^S$ and $\mathcal{F}^D$; binding energy $E$ and $D-$state probability $P^D$ as a function of $\nhid$. Columns 2-5 (6-9) provide results for sigmoid (softplus) activation functions.
 For completeness, we provide the benchmark exact values in the bottom row. 
 }
 \begin{center}
\begin{tabular}{| c | c c c c | c c c c |} 
\hline
%&  & Sigmoid & & & & Softplus & & \\
&  \multicolumn{4}{c|}{Sigmoid}  &  \multicolumn{4}{c|}{Softplus} \\ 
%$\nhid$ &$\mathcal{F}^{S}$ & $\mathcal{F}^{D}$ & $E_\textrm{ANN}$ (MeV) & $E_\textrm{GS}$ (MeV) & $P^D_\textrm{ANN}$ (\%) & $P^D_\textrm{GS}$ (\%) \\[0.5ex]  \hline
\hline
$\nhid$ & 
$\mathcal{F}^{S}$ & $\mathcal{F}^{D}$ & $E$ (MeV) & $P^D$ (\%) & 
$\mathcal{F}^{S}$ & $\mathcal{F}^{D}$ & $E$ (MeV) & $P^D$ (\%)\\[0.5ex]  \hline
% 2 & 0.92 (19) & 0.91 (19) & -0.79 (16) & 0.05 (1) & 0.92 (19) & 0.90 (19) & -0.79 (16) & 0.05 (1) \\
%  4 & 0.998290 (0.002539) & 0.994788 (0.010692) & -2.1483 (0.1587) & 0.0464 (0.0022) & 0.998008 (0.001558) & 0.993920 (0.021317) & -2.1478 (0.1186) & 0.0457 (0.0044) \\
  4 & 0.998(2) & 0.995(10) & -2.15(16) & 4.64(22) & 0.9980(16) & 0.993(21) & -2.14(12) & 4.57(44) \\
  6 & 0.99994(2) & 0.99995(4) & -2.223(1) & 4.51(7) & 0.99983(13) & 0.99976(20) & -2.220(6) & 4.50(12) \\
 %My values
 %  4 & 0.998(3) & 0.995(11) & -2.15(16) & 4.64(22) & 0.9980(16) & 0.994(21) & -2.15(12) & 4.57 (44) \\
 % 6 & 0.99994(2)& 0.99995(4) & -2.223(1) & 4.51(7) & 0.99983(13) & 0.99977(20) & -2.220(6) & 4.50 (12) \\
  8 & 0.999973(7) & 0.999974(22) & -2.2247(7) & 4.52(6) & 0.999950(21) & 0.999963(30) & -2.2243(18) & 4.52(9) \\
 10 & 0.999981(5) & 0.999974(21) & -2.2249(8) & 4.53(6) & 0.999964(7) & 0.999981(10) & -2.2248(8) & 4.50(7) \\
 12 & 0.999985(4) & 0.999983(20) & -2.2253(7) & 4.52(5) & 0.999970(6) & 0.999983(12) & -2.2251(7) & 4.51(8) \\
 14 & 0.999987(3) & 0.999986(14) & -2.2254(6) & 4.52(5) & 0.999973(4) & 0.999981(14) & -2.2251(7) & 4.51(8) \\
 16 & 0.999989(4) & 0.999986(14) & -2.2254(6) & 4.51(4) & 0.999975(6) & 0.999985(11) & -2.2252(8) & 4.53(6) \\
 18 & 0.999990(3) & 0.999982(17) & -2.2255(5) & 4.52(5) & 0.999975(4) & 0.999981(16) & -2.2251(8) & 4.52(7) \\
 20 & 0.999992(3) & 0.999985(11) & -2.2256(5) & 4.51(5) & 0.999976(5) & 0.999984(14) & -2.2252(8) & 4.51(7) \\ \hline 
 Exact & 1 & 1 & -2.2267 & 4.51 & 1 & 1 & -2.2267 & 4.51 \\
 \hline
\end{tabular}
\end{center}
\end{table*}

When a ground-state ANN wavefunction has been obtained, we quantify its quality by comparing it to the benchmark wavefunction from exact diagonalisation using a partial-wave fidelity, $\mathcal{F}^L$. This is akin to the overlap defined in Eq.~(\ref{eq:overlap}) with the replacement $ \psi_\textrm{targ}^L \to  \psi_\textrm{GS}^L$ \cite{Saito2018}. The closer $\mathcal{F}^L$ is to one, the closer our wavefunction reproduces the exact diagonalisation results. 

%\begin{table*}
% \caption{\label{table}
% VANN results for the overlaps $\mathcal{F}^S$ and $\mathcal{F}^D$ (columns 1 and 2); binding energy $E_\textrm{ANN}$ (3) and $D-$state probability $P^D_\textrm{ANN}$ (5) as a function of $\nhid$. For completeness, we provide the benchmark binding energy, $E_\textrm{GS}$, and probability, $P^D_\textrm{GS}$, in columns 5 and 7, respectively. 
% }
% \begin{center}
%\begin{tabular}{c c c c c c c} 
%\hline
%$\nhid$ &$\mathcal{F}^{S}$ & $\mathcal{F}^{D}$ & $E_\textrm{ANN}$ (MeV) & $E_\textrm{GS}$ (MeV) & $P^D_\textrm{ANN}$ (\%) & $P^D_\textrm{GS}$ (\%) \\[0.5ex]  \hline
% 2 & 0.96656(55) & 0.9454(22) & -0.8303(29) & -2.2267 & 5.66(49) & 4.51 \\
% 4 & 0.9988(10) & 0.99854(47) & -2.1963(197) & -2.2267 & 4.43(22) & 4.51 \\ 
% 6 & 0.999873(37) & 0.999867(39) & -2.2212(28) & -2.2267 & 4.49(67) & 4.51 \\ 
% 8 & 0.999940(21) & 0.999937(38) & -2.2240(12) & -2.2267 & 4.56(9) & 4.51 \\
% 10 & 0.999962(9) & 0.999981(13) & -2.2246(12) & -2.2267 & 4.55(8) & 4.51 \\ 
% 12 & 0.999967(9) & 0.999975(19) & -2.2246(11) & -2.2267 & 4.52(8) & 4.51 \\
% 14 & 0.999968(12) & 0.999973(15) & -2.2247(12) & -2.2267 & 4.49(8) & 4.51 \\ 
% 16 & 0.999975(5) & 0.999983(10) & -2.2252(7) & -2.2267 & 4.55(7) & 4.51 \\
% 18 & 0.999974(8) & 0.999987(5) & -2.2250(11) & -2.2267 & 4.51(6) & 4.51 \\
% 20 & 0.999975(7) & 0.999983(12) & -2.2249(9) & -2.2267 & 4.48(6) & 4.51 \\[1ex]
% \hline\hline
%\end{tabular}
%\end{center}
%\end{table*}

The main results of this paper are reported in Fig.~\ref{fig:energyNhid} and, in a tabular form, in Table~\ref{table}. %We exclude the results of the $\nhid=2$ runs from the Figure for clarity, but show them in the Table. 
%Results up to $\nhid=100$ are discussed in~\ref{app:variance}. 
In all cases, we report outcomes obtained for both sigmoid and softplus activation functions. 
%We take the fact that the results obtained with both functions are similar as an indication that the ML approach is robust. 
%
With an $\nhid=2$ model, not shown for brevity, the deuteron is already bound by $\approx 0.8$ MeV.
For $\nhid=4$, the quality of the ANN ansatz is relatively competitive, with fidelities within $2 \%$ of $\mathcal{F}^L=1$,  and a binding energy that is already within about $\approx 5\%$ of the benchmark value, albeit with a significant standard deviation. At the level of $\nhid=6$, we already obtain energies (fidelities) that are accurate within $10$ keV ($0.05 \%$).
As $\nhid$ increases, the energy approaches the benchmark, and stabilises around $\nhid \approx 10$. Above this value, we find a relative agreement of the order of $\approx 2$ keV in energies. 
The error in fidelities remains relatively constant above $\nhid \approx 10$ too, at a level of $\approx  0.005 \%$ across all the models.

Having access to the wavefunctions, we can also compute structural properties of the deuteron. The $D-$state probability, $P^D_\textrm{ANN}=\left \langle \psi_\textrm{ANN}^D \right | \left. \psi_\textrm{ANN}^D \right \rangle$, is correlated with the strength of the tensor force. With as little as $4$ hidden nodes the admixture between the $S$ and $D$ states is off by just over $0.1\%$. 
As $\nhid$ increases, the values approach the benchmark $P^D_\textrm{GS}=4.51 \%$. 
The bottom panel of Fig.~\ref{fig:energyNhid} indicates that the network is able to predict the admixture between the $S$ and $D$ state with a variance of less than $0.1 \%$. 
%Here, again, we do not find a significant rise in variance as $\nhid$ increases.

%In fact, the variance obtained from the $50-$run standard deviation remains relatively constant above $\nhid \approx 10$. We postulate that the variance does not in our model because the integrals associated to the (global) energy cost function and to fideltiies carry a $q^2$ prefactor, as shown in Eq.~(\ref{eq:overlap}). This in turn means that there is no penalty associated to changing the low-momentum regions of the wavefunction. We provide an extended discussion about this constant variance up to $\nhid=100$ in~\ref{app:variance}. 

When it comes to different activation functions, the sigmoid and softplus results provide qualitatively similar results. We take this is as a sign of robustness in the methodology. At a quantitative level, the sigmoid calculations outperform the softplus results. Sigmoids seem to provide results that are closer to benchmarks and have relatively smaller variances. As seen in the Table~\ref{table}, but also in the $\nhid$ convergence shown in the central panels of Fig.~\ref{fig:energyNhid}, the fidelities predicted by the sigmoid ANN for the $S-$state are substantially better than those predicted by the softplus ANN. $D-$state fidelities, in contrast, have a similar level of quality for both activation functions. 

Variational calculations with the same N3LO interaction typically require $\approx 8$ parameters to find energies within $0.1$ MeV of the exact value \cite{Bogner2006,Bogner2006b,Anderson2010}. We are not aware of other variational calculations in momentum space that use more parameters. We have however set up a stochastic variational method solution to the deuteron, with exactly the same momentum-space set-up \cite{Rios2020}. We find that, to get an accuracy equivalent to the $\nhid=6$ case of the ANN models, $24-32$ parameters are required. We take this as an indication that other variational methods require a similar number of parameters to reach the same level of agreement with exact benchmarks.

\begin{figure}
\begin{center}
\includegraphics[width=\linewidth]{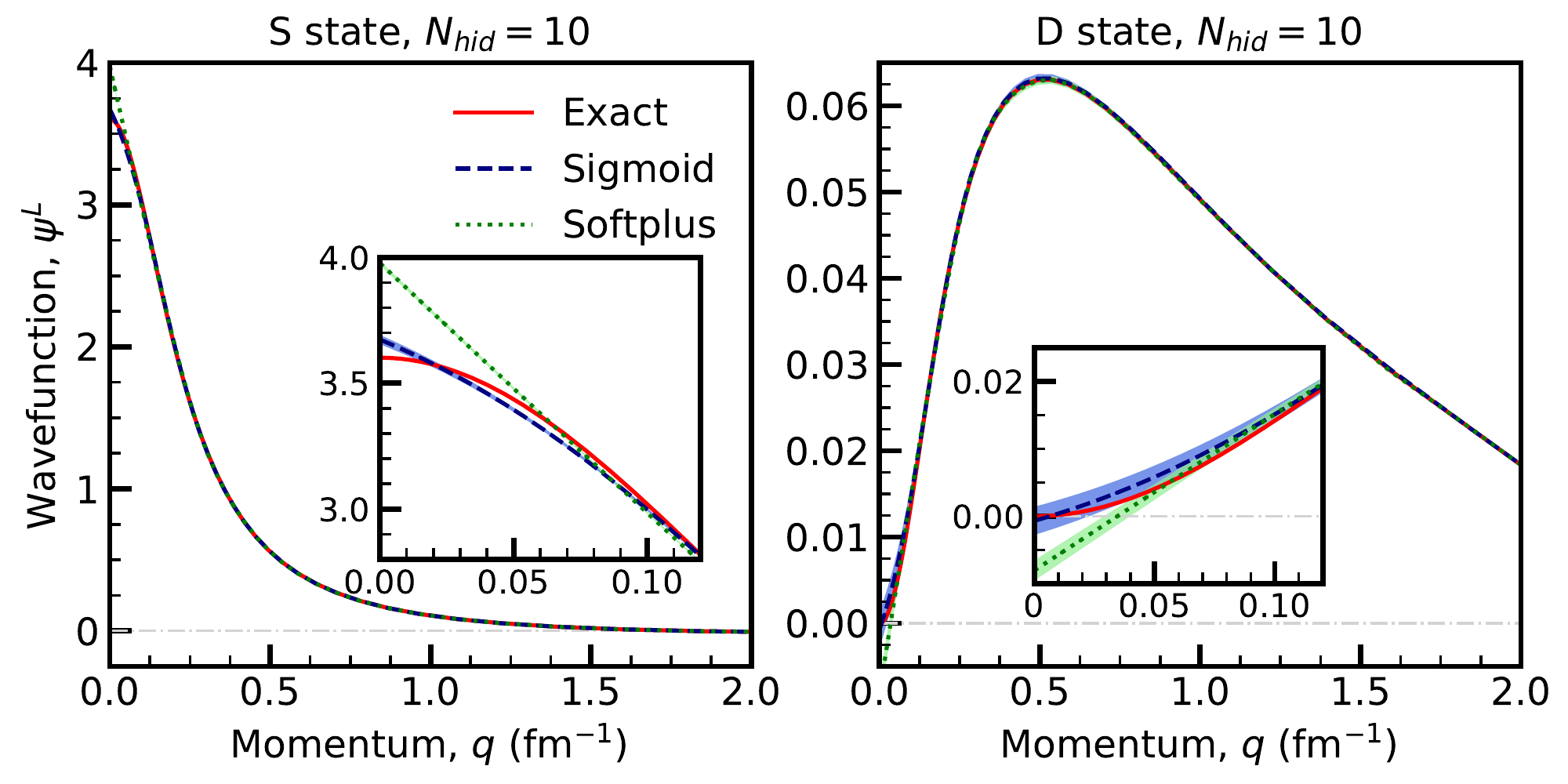}
\caption{\label{fig:comparewfs} Left (right) panel: the $S$ ($D$) state wavefunction as a function of momentum. Exact wavefunctions (solid lines) are compared to the  $\nhid=10$ ANN wavefunctions using sigmoid (dashed) and softplus (dotted) activation functions. The bands correspond to the standard deviation associated to $50$ different initialisation runs. 
%, but it is difficult to see in the scale of the figure. 
} 
\end{center}
\end{figure}

Despite a relatively low variance and a very small error in the fidelities, the energy associated to the VANN wavefunction never quite reaches the benchmark value as $\nhid$ increases. In an attempt to understand the origin of the discrepancy, we compare in
Fig.~\ref{fig:comparewfs} the exact wavefunctions (solid lines) and the $\nhid=10$ sigmoid (dashed) and softplus (dotted) ANN predictions for the $S$ (left panel) and $D$ states (right).
The width associated to the $50-$run standard deviation is included in the ANN wavefunctions, but it is hard to see on this scale.
The agreement between ANN and exact wavefunctions is excellent across a wide range of momenta, including the change of sign of the $S-$state wavefunction around $q=1.8$ fm$^{-1}$. The only region where a significant discrepancy is visible is close to the origin, $q < 0.05$ fm$^{-1}$. There, the softplus ANN overshoots linearly the $S-$state wavefunction, and undershoots the $D-$state result. While the sigmoid predictions have some inherent curvature, they still miss the quantitative dependence of $\psi_\textrm{GS}$ at low momentum. 

The low-momentum mismatch is to a certain extent expected. All the integrals, including those associated to the energy cost function, carry a $q^2$ factor [see Eq.~(\ref{eq:overlap})]. Consequently, there is no energetic penalty for the VANN energy to miss the correct shape at zero momentum. 
As discussed further in~\ref{app:variance}, the $q^2$ phase-space factor is also largely responsible for the constant variance in all quantities for $\nhid>10$. Having said that, the presence of this factor also implies that energy differences with respect to the exact case must originate at finite momenta. 
Our preliminary analysis indicates that the small differences with respect to the benchmark energy value originate at relatively large momenta (in the region $2-10$ fm$^{-1}$). 
The correct low-momentum boundary conditions may need to be explicitly incorporated to further improve these energy predictions. We note however that the asymptotics in this region are $L-$dependent, $\psi^L(q) \approx q^L$. Including these or any further information explicitly in the ansatz would require additional layers in the ANN, beyond the minimal philosophy of our exploratory  analysis.

%As discussed further in~\ref{app:variance}, the $q^2$ phase-space factor in the global energy cost function can explain the discrepancy. This results in small mismatches in the energy, which the network cannot overcome without additional, local information in the wavefunction ansatz. For the unbounded softplus activation function, this seems to result in extrapolations towards the linear (large $x$) regime. In contrast, the bounded behavior of the sigmoid is more prone to display the saturation needed in the $S-$state wavefunction at the origin. 
%may result in arbitrarily large $\mathbfcal{W}^{(1)}$ weights which only explore the linear regime of the softplus function. 

%  going beyond both the exploratory scope and minimal nature of our work.

%In fact, the variance obtained from the $50-$run standard deviation remains relatively constant above $\nhid \approx 10$. We postulate that the variance does not in our model because the integrals associated to the (global) energy cost function and to fideltiies carry a $q^2$ prefactor, as shown in Eq.~(\ref{eq:overlap}). This in turn means that there is no penalty associated to changing the low-momentum regions of the wavefunction. We provide an extended discussion about this constant variance up to $\nhid=100$ in~\ref{app:variance}. 

\section{Conclusions}
\label{sec:conclusions}

Our results show, for the first time, that VANN techniques can be used successfully in solving bound-state nuclear physics problems. We find that minimal networks with a single layer and as little as $\nhid=6$ nodes provide faithful representations of the exact wavefunction, providing binding energies within a few keV of benchmarks - or $0.1 \%$ in relative value. 
%The agreement generally improves as $\nhid$ increases, but saturates around $\nhid \approx 10$. 
Structural properties, like $P_D$, are a by-product of the calculation and show similar levels of agreement. The variance of the models remain rather constant for a wide range of $\nhid$. We speculate that this constant variance, of the order of a fraction of a percent, arises as a consequence of the $q^2$ phase-space factors in all the integrals associated to physical values. 

For the deuteron, these results are not yet competitive in terms of computing time. Our results however indicate that very simple architectures with a small number of nodes are already good starting points, yielding accurate results. Our simple implementation using existing ML tools is effectively solving a one-body problem in relative coordinates in a fixed momentum mesh. It is not designed to solve fully fledged many-body systems. If the simplicity in the ANN ansatz could be exploited for heavier systems, the scaling in computing time of VANN techniques may remain relatively mild. If this is the case, one may be able to tackle heavier systems with this variational \emph{ab initio} approach, as already demonstrated in quantum chemistry \cite{Pfau2019,Hermann2019,Choo2020}. 

We foresee some bottlenecks before extending the reach of VANN techniques to higher mass numbers. 
First, techniques to explicitly include antisymmetrisation in the many-body wavefunction need to be developed. Recent results exploiting permutation-equivariant ANNs can provide a way forward \cite{Pfau2019}. Second, the network will have to deal with several configurations, as well as two- and three-nucleon interactions. 
%This first application has already shown that VANNs can deal with more than one nuclear state at once. 
Third, and more important, it remains to be seen whether a generic extension of VANNs to incorporate arbitrary spin and isospin is possible. 
This may require specifically tailored deep ANN architectures. Only after these issues have been tackled, it will become clear whether ML is a competitive tool for \emph{ab initio} nuclear physics. 

\section{Acknowledgments}

This work is supported by the UK Science and Technology Facilities Council (STFC) through grant ST/P005314/1. We thank Pierre Arthuis and Mehdi Drissi for a careful reading of the manuscript and for useful discussions. 

\appendix

%\bibliographystyle{iopart-num}
%\bibliography{biblio}

\section{RMSprop}
\label{app:RMSprop}

We use the Root Mean Square Propagation (RMSprop) method in all the minimisation processes involved in our work \cite{Mehta2019,Hinton2012}. This deterministic approach is relatively popular in the ML community and can be thought of as an extension of the standard gradient descent method, including additional information on the second moment of the gradient. 
In a standard gradient descent scenario, ANN weights, $\mathcal{W}_t$, are updated at each optimisation iteration, $t$, following the direction of maximum change in the cost function $\mathcal{C}$, 
 \begin{align}
     \mathcal{W}_{t+1} = \mathcal{W}_{t} - \alpha\frac{\partial \mathcal{C}}{\partial \mathcal{W}_{t}} \, .
 \end{align}
 The hyperparameter $\alpha$ is generally referred to as learning rate.
In contrast, in the RMSprop algorithm, the updates proceed in two steps,
 \begin{align}
     \mathcal{V}_{t+1} &= \beta \mathcal{V}_{t} + (1-\beta)\left(\frac{\partial \mathcal{C}}{\partial \mathcal{W}_{t}}\right)^{2} \, , 
     \label{eq:rmsprop1}\\
     \mathcal{W}_{t+1} &= \mathcal{W}_{t} - \frac{\alpha}{\sqrt{\mathcal{V}_{t}}+\epsilon}\frac{\partial \mathcal{C}}{\partial \mathcal{W}_{t}} \, .
     \label{eq:rmsprop2}
 \end{align}
In the first step, $\mathcal{V}$ provides an exponential moving average (EMA) of the square of the gradient in the direction of a particular weight within the network. The starting point is $\mathcal{V}_{t=0}=0$. The smoothing hyperparameter, $\beta \in [0,1)$, controls the importance of the history of the square of the gradient.
The EMA allows the recent history of the gradient to be stored efficiently. The second term, Eq.~(\ref{eq:rmsprop2}), is akin to the standard gradient descent, but the prefactor $\sqrt{\mathcal{V}_{t}}+\epsilon$ regulates the learning rate, $\alpha$. In this process, all weights have a learning rate that is modified to better suit the local geometry in the cost function minimisation landscape.
The regularisation hyperparameter $\epsilon$ has a small value to stop any divide-by-zero errors in the event that $\mathcal{V}_{t}=0$. 
In our implementation, we use a learning rate $\alpha = 10^{-2}$, a smoothing constant $\beta = 0.9$, and a numerical stability constant $\epsilon = 10^{-8}$. We note that PyTorch's implementation of RMSprop differs from Tensorflow in the prefactor of Eq.~(\ref{eq:rmsprop2}), where the numerical stability constant $\epsilon$ is included within the square root. 

RMSprop requires access to the explicit derivatives of the cost function with respect to the weights, $\partial \mathcal{C} / \partial \mathcal{W}_{t}$. 
These are calculated via PyTorch's \verb+autograd+ library. \verb+autograd+ is a form of differentiation which is not based on numerical nor symbolic methods \cite{baydin2017automatic}.% This automatic differentiation tool considers all the chain-rule steps necessary in taking derivatives with respect to any weight. Ultimately, it uses pre-computed derivatives of activation functions as the initial state in a ``backpropagation" of sorts for the derivatives. 
%This allows for better scaling for networks with large number of parameters, no round-off or truncation errors from numerical methods, no cases of `expression swell' from symbolic methods and also it can make use of control flow procedures like branching, loops, and recursion \cite{baydin2017automatic}. 
The \verb+autograd+ library supports reverse-mode automatic differentiation \cite{PyTorch,NEURIPS2019_9015}. This calculates the gradient of the network with respect to a given parameter by exploiting a tree map describing the dependencies of all nodes on different variables. \verb+autograd+ requires pre-computed derivatives at each node, and subsequently exploits the chain-rule to calculate derivatives throughout the network. %This calculates the gradient of the network with respect to a given parameter by exploiting a tree map describing the dependencies of all nodes on different variables, and a subsequent implementation of the chain-rule. %Firstly, a tree is built to describe the variable dependences in the network. In this forward phase, \verb+autograd+ identifies the network input, $\left(q\right)$,  is passed through the network to calculate its associated output $\left(\psi^S\left(k\right), \psi^D\left(k\right)\right)$ as well as the intermediate hidden values in the hidden layer. Next, in the reverse phase, the derivatives are calculated by propagating the local derivatives of intermediate values in reverse from the output to the input by use of the chain-rule. This allows for the exact derivative to be calculated - 
A pedagogical example of the use of \verb+autograd+ can be found in section 3.2 of Ref.~\cite{baydin2017automatic}.
Finally, we note that we do not need to perform any derivatives of the wavefunction itself as a function of momentum $q$. This is in contrast to real-space implementations, where gradients as a function of spatial coordinates are required to compute many-body kinetic energies.

\section{Wavefunction variance analysis}
\label{app:variance}

\begin{figure}
\begin{center}
\includegraphics[width=\linewidth]{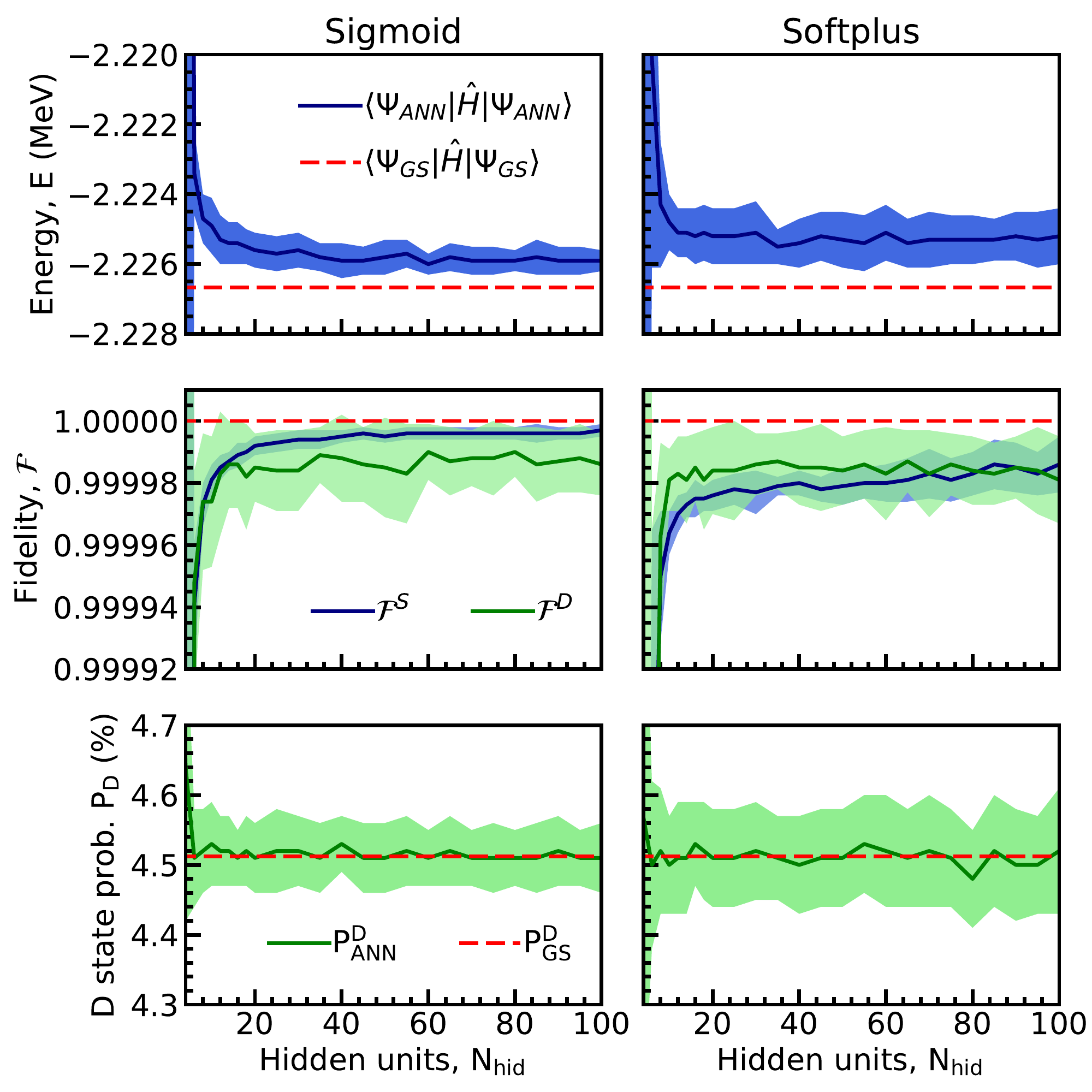}
\caption{\label{fig:6panels_app} 
Same as Fig.~\ref{fig:energyNhid} but with an extended range in $\nhid$. Note the change in scales in the $y-$axis, which are significantly closer to the benchmark values here.
} 
\end{center}
\end{figure}

In a typical bias-variance tradeoff scenario, one expects the variance on the ANN predictions to initially decrease with $\nhid$, as the model improves its flexibility, only to see it increase above an optimal value, $\nhid^\textrm{opt}$, as overfitting takes over \cite{Mehta2019}. 
Instead, we find that the variance obtained as the standard deviation of $50$ VANN runs remains relatively constant above $\nhid \approx 10$. We show this graphically in Fig.~\ref{fig:6panels_app}, where the $\nhid$ range of Fig.~\ref{fig:energyNhid} is extended up to $\nhid=100$. The figure clearly indicates that the variance remains constant across the whole $\nhid$ range. In addition, the quality of the results in terms of difference with respect to benchmarks also saturates above a given threshold value. 

%We postulate that both of these issues (difference with respect to benchmark and constant variance) are the result of the same shortcoming of our implementation. 
We postulate that the constant variance is the result of a shortcoming of our implementation - namely the fact that we work with relative momenta in spherical coordinates. As a consequence, all the integrals associated to physical quantities carry a $q^2$ prefactor, as shown in Eq.~(\ref{eq:overlap}). This is the case for the (global) energy cost function, too. In other words, there is no penalty associated to changing the zero (or, for that matter, the low-momentum values) of the wavefunction. In principle, the wavefunction could be arbitrarily far away from the benchmark, without additional costs. In practice, however, the continuity of the activation function and its asymptotic properties at large values of input are reflected in this region. We stress that the local variability at low $q$ would not be identified in any of the global, integrated physical measures, like the energy, the fidelity or the $D-$state probability.

\begin{figure}
\includegraphics[width=\linewidth]{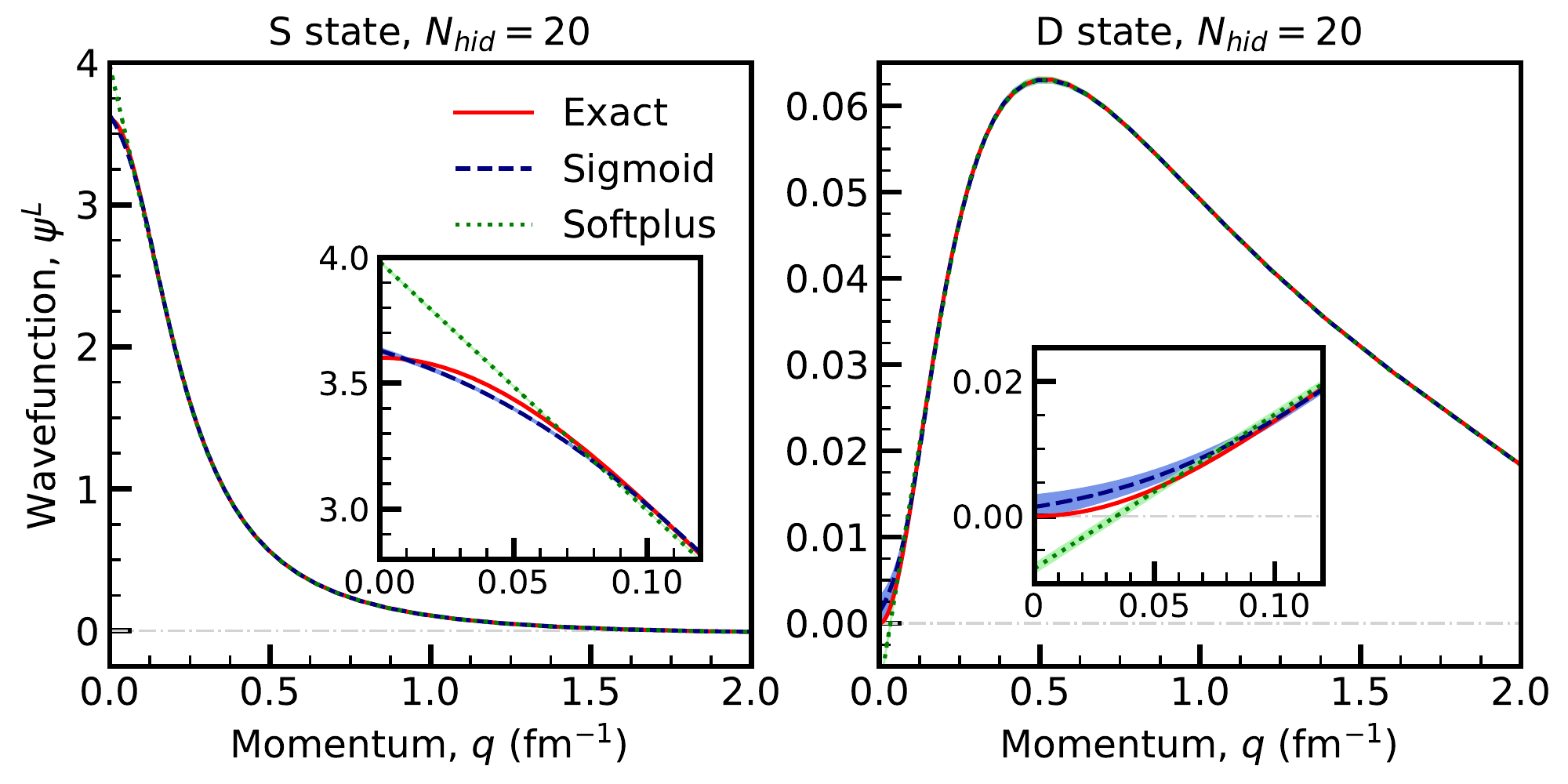}
\includegraphics[width=\linewidth]{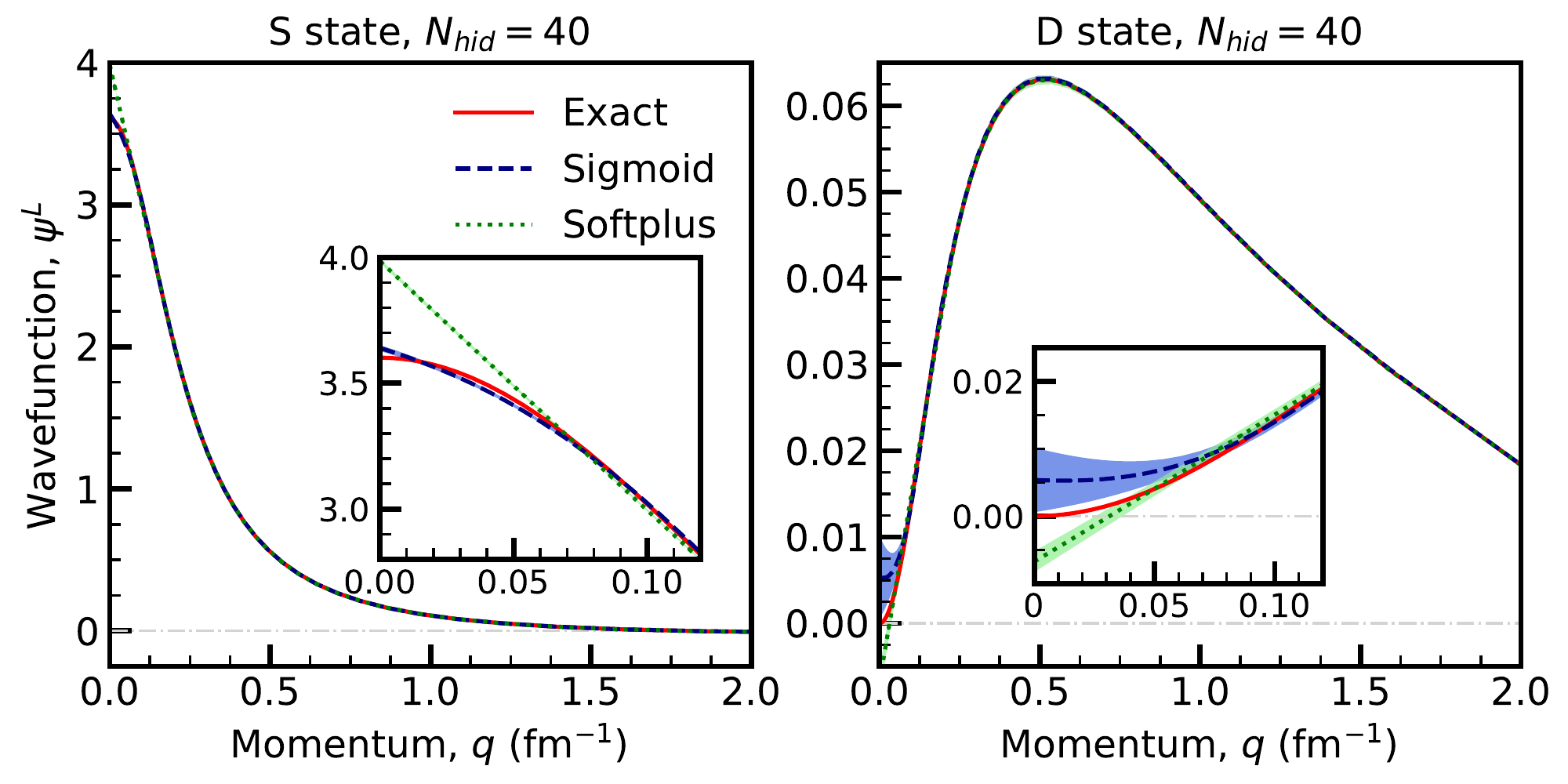}
\includegraphics[width=\linewidth]{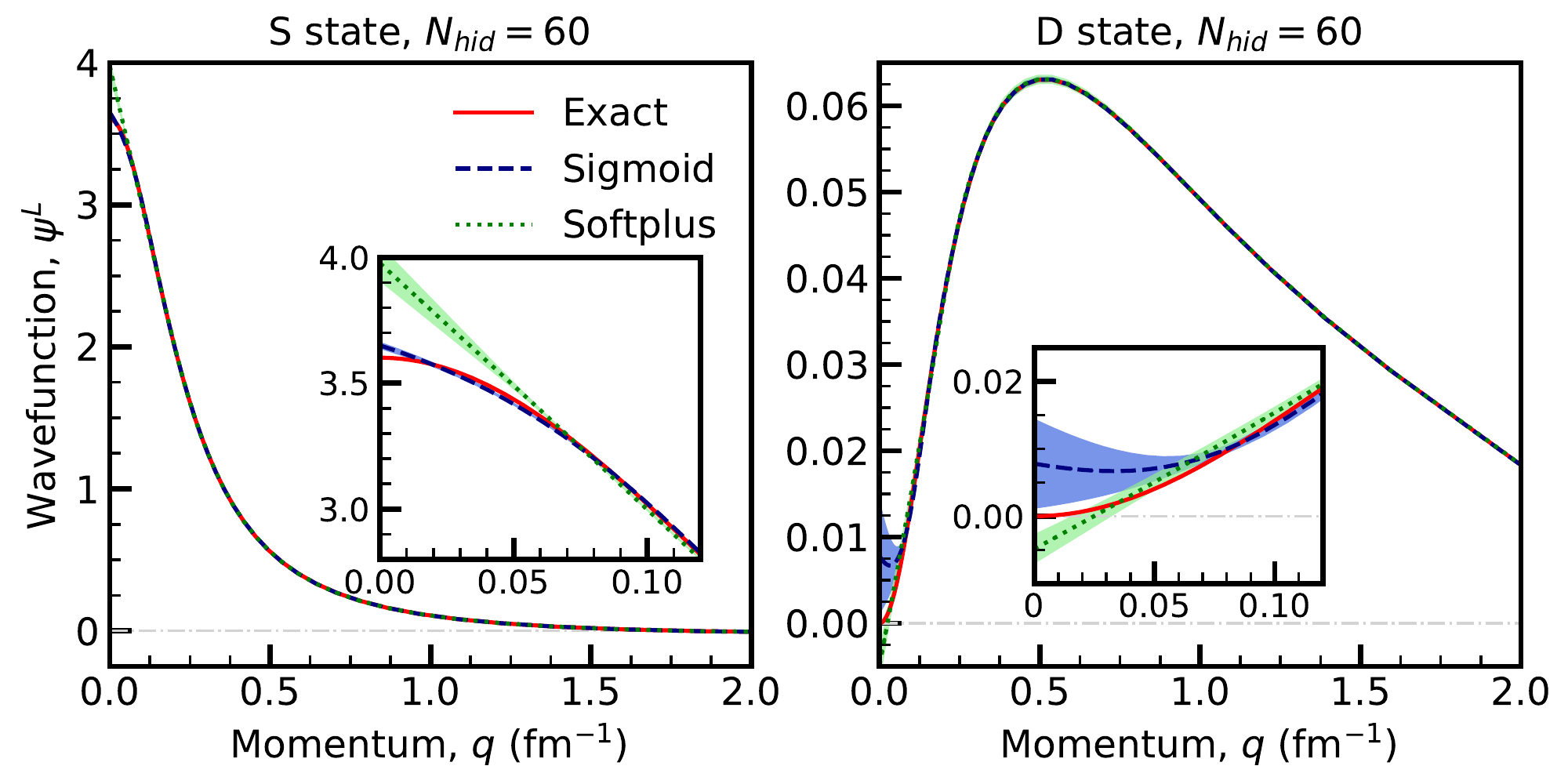}
\includegraphics[width=\linewidth]{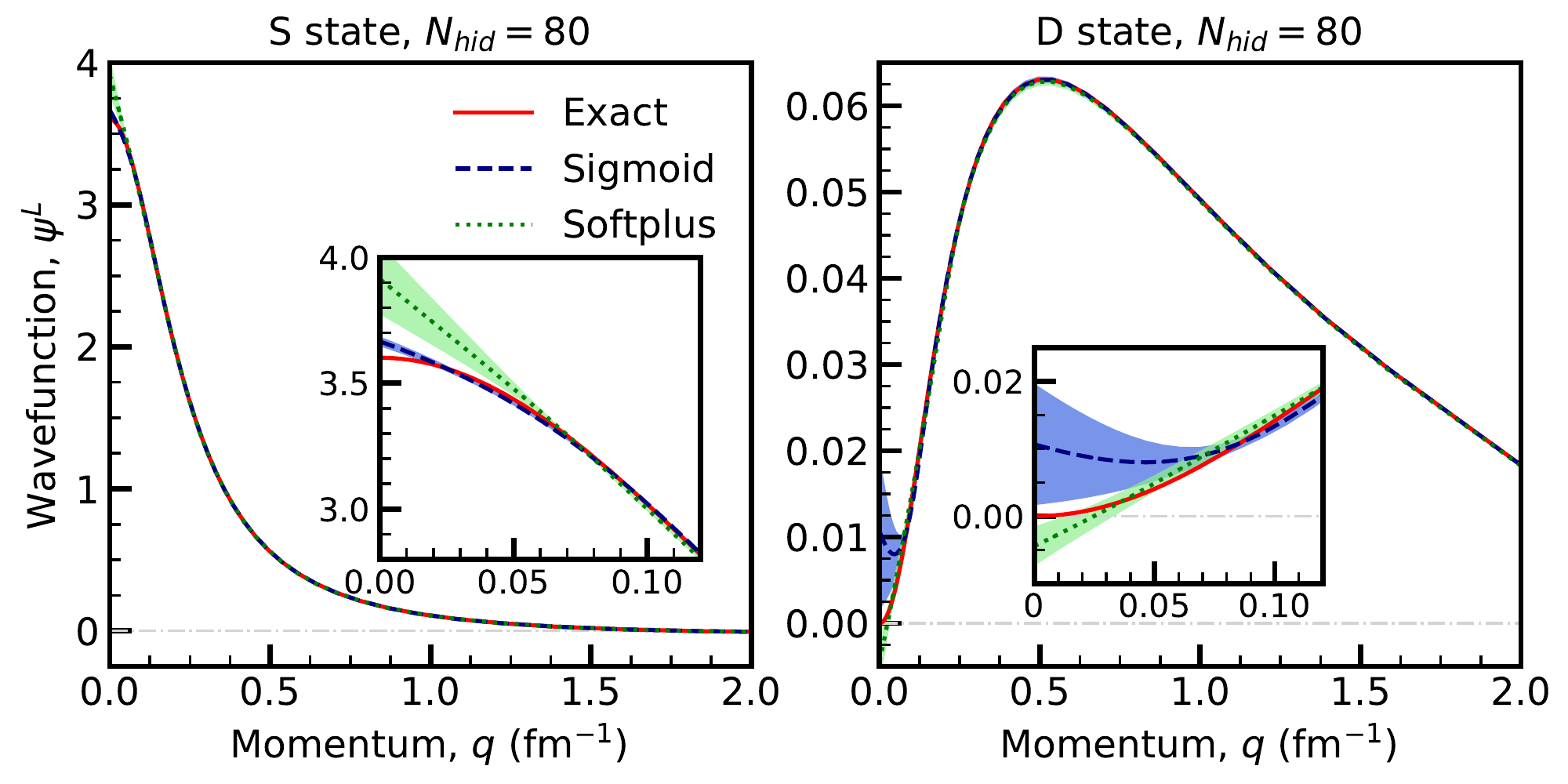}
\includegraphics[width=\linewidth]{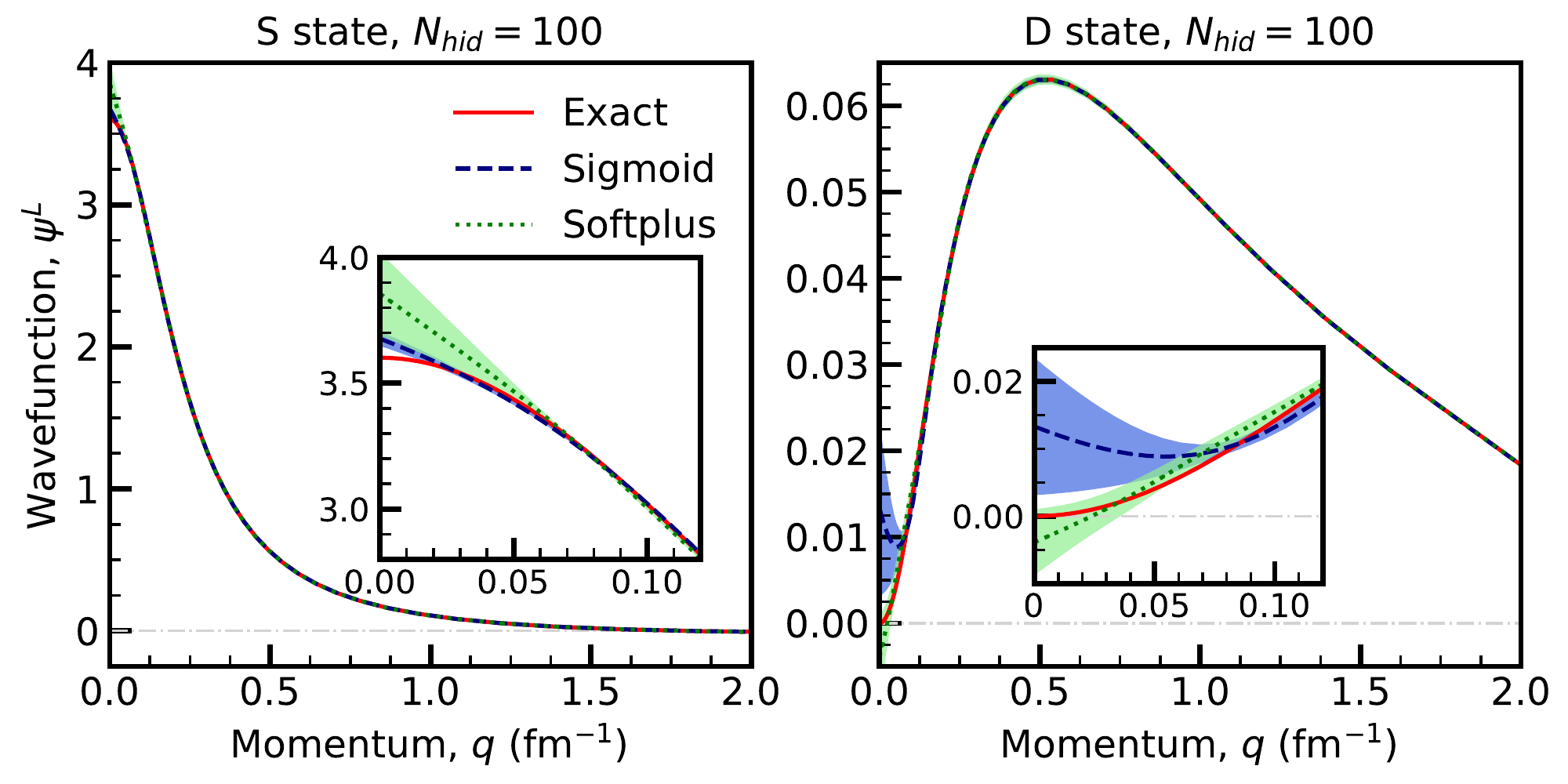}
\caption{ \label{fig:comparison_nhid}
Same as Fig.~\ref{fig:comparewfs}, but for increasing values of $\nhid$ from $\nhid=20$ (top row) to $\nhid=100$ (bottom row).
}
\end{figure}

We provide proof of this behaviour in Fig.~\ref{fig:comparison_nhid}, where we show the equivalent to Fig.~\ref{fig:comparewfs} for a range of values of $\nhid$. The top panel corresponds to $\nhid=20$, and $\nhid$ increases towards the bottom, which shows the extreme case of $\nhid=100$. The wavefunction above $q \approx 0.10$ fm$^{-1}$ is reproduced by both the sigmoid and softplus ans\"atze to the wavefunctions. Towards the origin, however, both trial wavefunction struggle to reproduce the correct asymptotics. 
The low-momentum ANN predictions with sigmoid activation functions are much closer to the exact $S-$state wavefunctions than the corresponding softplus ANN.
For the $D-$state, the softplus ANN generally misses the low-momentum asymptotics and undershoots the wavefunction linearly. The centroid of the sigmoid also misses the boundary condition at the origin, and in fact shows an increase in curvature as $\nhid$ grows. These different behaviours towards the origin seem to reflect the bounded (sigmoid) or unbounded (softplus) nature of the activation functions at large values of input. 

%The linear behaviour of the softplus function is characteristic of its large-input behavior, as is the saturating behaviour of the sigmoid. 

%Whereas softplus overshoots the origin values linearly, the sigmoid has a hard time reaching the origin of the $D-$state. 

Our arguments relate to the size of the bands towards the origin, shown in the insets. These figures demonstrate that the variance in the low-momentum values of the wavefunction increases with $\nhid$. As opposed to global integrated measures, local regions of the wavefunction are  subject to a bias-variance trade-off. 
We take this as an indication that, above a certain threshold value of $\nhid^\textrm{opt}$, an increase in ANN complexity does not bring in an increase in wavefunction quality. 
%As opposed to the $S-$state, the sigmoid ANN produces a larger variance compared to the softplus predictions. 

\begin{figure}
\begin{center}
\includegraphics[width=\linewidth]{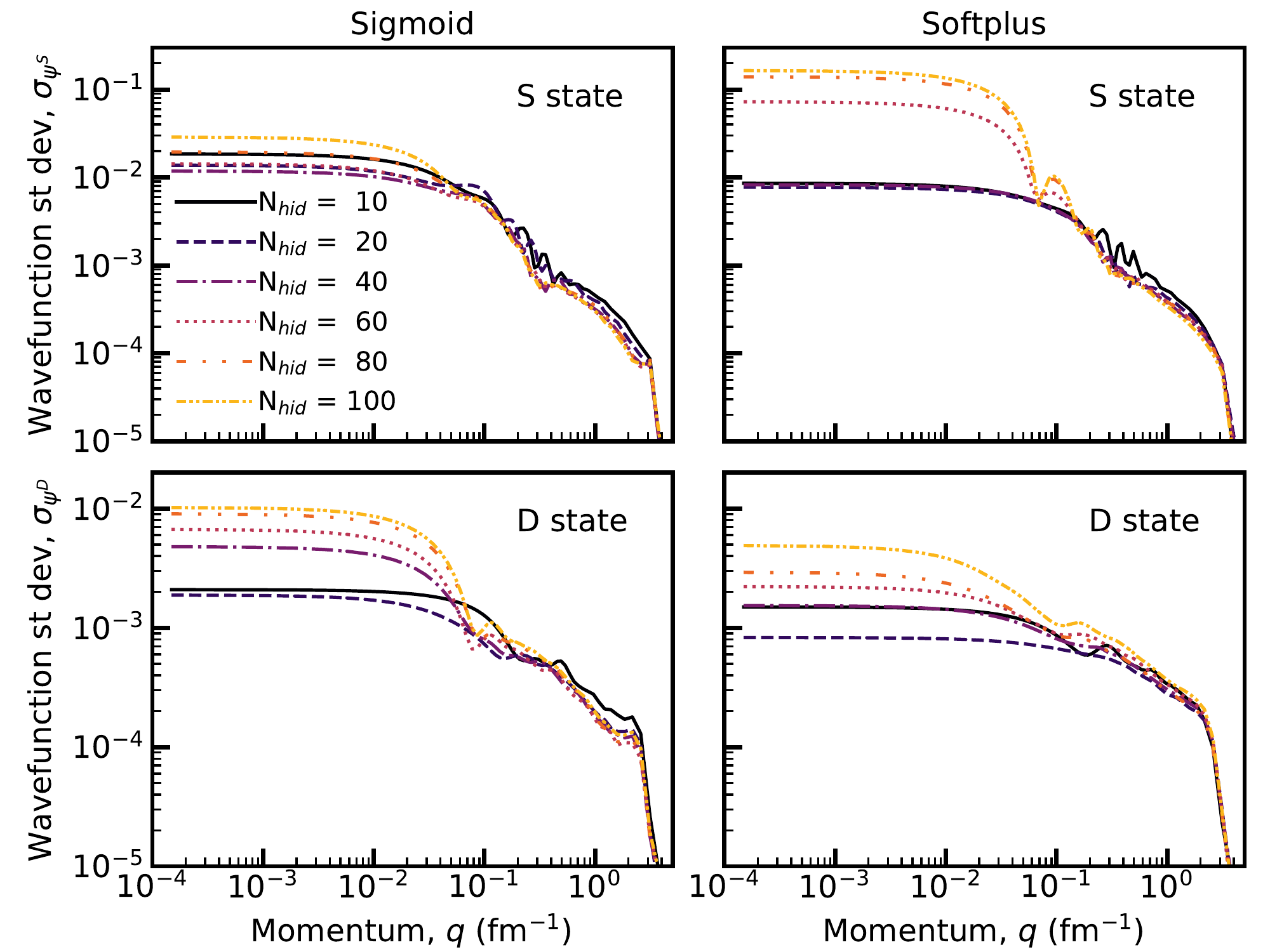}
\caption{\label{fig:std} 
Top panels: standard deviation of the $S-$state wavefunction obtained after $50$ minimisation runs as a function of momentum for the sigmoid (left) and softplus (right) activation functions. Different lines correspond to different values of $\nhid$. Bottom panels: the same for the $D-$state.
} 
\end{center}
\end{figure}

More details are provided in Fig.~\ref{fig:std}. Rather than showing the wavefunction itself, we focus here on the standard deviation of the wavefunction,  $\sigma_{\psi^L}$, i.e. the width of the bands in Fig.~\ref{fig:comparison_nhid}. This is shown as a function of momentum in a log-log scale, to magnify the differences. Left (right) panels correspond to sigmoid (softplus) activation functions, and top (bottom) panels show results for the $S-$ ($D-$)state. Different lines correspond to different values of $\nhid$. 
First, we reiterate the message that the variance of the wavefunction is maximal at the lowest momenta. In fact, the variance decreases sharply above $q \approx 1$ fm$^{-1}$.
Second, the $\nhid$ dependence is also rather informative, as it indicates that the minimal variance in all the models is reached around $\nhid^\textrm{opt} \approx 20$. Values of $\nhid$ below or above the optimum value provide larger variances in  wavefunctions. For the optimal value in the Figure, the differences between the underlying activation functions are small for the $S-$state, and within a factor of $2$ for the $D-$state. 

Finally, the dependence of $\sigma_{\psi^L}$ on $\nhid$ is also prone to relatively large jumps, as seen in the softplus $S-$state results between $\nhid=40$ and $60$ or in the sigmoid $D-$state predictions between $\nhid=20$ and $40$. A more detailed analysis may be needed to fully understand the origin, shape and $\nhid$ dependence of these structures. Alternatively, it may be more interesting to incorporate the knowledge of low-momentum asymptotics in the ANN ansatz. In real-space implementations for electronic structure, the exponentially decaying asymptotics at large distances and the corresponding cusp conditions are known, and deliberately coded into the ANN as an additional layer \cite{Pfau2019,Hermann2019}. In the case of the deuteron, the implementation of the $L-$dependent boundary conditions would require an additional layer in the network to match wavefunctions into analytical behaviours at low values of momentum. We leave the analysis of these types of extensions for future work. 

%\bibliographystyleApp{iopart-num}
%\bibliographyApp{biblio_app}

\bibliographystyle{iopart-num}
\bibliography{biblio}

\end{document}